\documentclass{article}
\usepackage[english]{babel}
\usepackage{amsfonts}
\usepackage{amsmath}
\usepackage{amssymb}
\usepackage{amsthm}
\usepackage{array}
\usepackage{multirow}
\usepackage{float}
\usepackage{booktabs}
\usepackage{color}
\usepackage{csquotes}

\usepackage{graphicx}
\usepackage{hyperref}

\def\join{\mathop\vee}
\def\meet{\mathop\wedge}
\def\eA{\textbf{A}}
\def\eB{\textbf{B}}
\def\eC{\textbf{C}}
\def\eD{\textbf{D}}
\def\eI{\textbf{I}}
\def\eX{\textbf{X}}
\def\eY{\textbf{Y}}
\def\ea{\textbf{a}}
\def\eb{\textbf{b}}
\def\ec{\textbf{c}}
\def\ed{\textbf{d}}
\def\ef{\textbf{f}}
\def\eg{\textbf{g}}
\def\ei{\textbf{i}}
\def\et{\textbf{t}}
\def\eu{\textbf{u}}
\def\ev{\textbf{v}}
\def\ew{\textbf{w}}
\def\ex{\textbf{x}}
\def\ey{\textbf{y}}
\def\ez{\textbf{z}}

\def\Pr{\mbox{Pr}}
\def\lor{\;\mbox{or}\;}
\def\land{\;\mbox{and}\;}

\def\<{\triangleleft}

\DeclareRobustCommand{\concatbase}{\mathbin{\rotatebox[origin=c]{90}{\scalebox{.7}{(\kern1ex)}}}}

\def\join{\vee}
\def\meet{\wedge}
\def\||{\,||\,}

\newcommand{\red}{\color{black}}

\newcommand{\black}{\color{black}}

\newtheorem{theorem}{Theorem}
\newtheorem{definition}{Definition}

\begin{document}
\title{Lattices and Their Consistent Quantification \thanks{K. H. Knuth, 2018. Lattices and their consistent quantification, \textit{Annalen der Physik}, 1700370. \url{https://doi.org/10.1002/andp.201700370}}}
\author{Kevin H. Knuth\\Department of Physics\\University at Albany, Albany NY USA}

\maketitle

\abstract{
  This paper introduces the order-theoretic concept of lattices along with the concept of consistent quantification where lattice elements are mapped to real numbers in such a way that preserves some aspect of the order-theoretic structure.  Symmetries, such as associativity, constrain consistent quantification, and lead to a constraint equation known as the sum rule.  Distributivity in distributive lattices also constrains consistent quantification and leads to a product rule.  The sum and product rules, which are familiar from, but not unique to, probability theory, arise from the fact that logical statements form a distributive (Boolean) lattice, which exhibits the requisite symmetries.
}


\section{Introduction}

In science, especially theoretical physics, it is critical that we understand precisely why our successful theories work.  Why our theories are the way they are.  Unfortunately, this is not always obvious.  In fact, the situation is possibly more dire in that it has not been precisely clear why mathematics should be of any use at all in describing the physical world in the first place.  This issue was best highlighted in Wigner's essay ``The Unreasonable Effectiveness of Mathematics in the Natural Sciences'' \cite{Wigner:1960}, which was followed two decades later by a related essay by Hamming \cite{Hamming:1980}.  In his essay, Hamming remarks upon the surprising utility of number \cite{Hamming:1980}:
\begin{displayquote}
``I have tried, with little success, to get some of my friends to understand my amazement that the abstraction of integers for counting is both possible and useful. Is it not remarkable that 6 sheep plus 7 sheep make 13 sheep; that 6 stones plus 7 stones make 13 stones? Is it not a miracle that the universe is so constructed that such a simple abstraction as a number is possible? To me this is one of the strongest examples of the unreasonable effectiveness of mathematics. Indeed, I find it both strange and unexplainable.''
\end{displayquote}

It is reasonable to ask why addition is almost universally applicable when we combine things \cite{Knuth:FQXI2015}.  A careless, but informed, respondent might claim that this has to do with measure theory.  However, measure theory is based on additivity being an axiom, which means that if we take measure theory as a foundation we are simply \emph{assuming} that additivity, which is a central component to our theories, holds.

This is an unacceptable state for our theories.  It ought to be of great benefit to understand \emph{why} we add numbers when we combine things.  Why does the resistance of two resistors in series sum?  Why does one have linear superposition of electric fields?  Why is the total energy of a system found by summing the energies of each subsystem?  In statistical mechanics, variables that depend on the quantity of stuff, i.e. variables that sum when subsystems are considered together, are called \emph{extensive variables}.  It should be of utility to understand why some variables are extensive, especially since there exists a relatively recent mass of work focused on non-extensive entropies \cite{Tsallis:1995, Tsallis:2009}, which has been disputed on foundational grounds \cite{Presse:2014}.

The ubiquity of additivity, and more specifically, the inclusion-exclusion relation \cite{Rota:combinatorics, Krishnamurthy:combinatorics, Klain&Rota}, is a clue that there is something deeper \cite{Knuth:FQXI2015} lying beneath the accepted foundation of measure theory.  Examples include, but are not limited to,\\
\emph{Probability Theory}
\begin{equation}
\Pr(A \lor B | C) = \Pr(A | C) + \Pr(B | C) - \Pr(A \land B | C),
\end{equation}
\emph{Information Theory} (mutual information)
\begin{equation}
MI(A ; B) = H(A) + H(B) - H(A, B),
\end{equation}
for which $MI$ is the mutual information and $H$ is the entropy, the \emph{relationships among integral divisors}
\begin{equation}
\log(\mbox{LCM}(A,B)) = \log(A) + \log(B) - \log(\mbox{GCD}(A,B)),
\end{equation}
for which $\mbox{LCM}$ is the least common multiple and $\mbox{GCM}$ is the greatest common divisor, and \\
\emph{quantum amplitudes in the three-slit problem} \cite{Sorkin:1994}
\begin{equation}
\begin{split}
I_3(A,B,C) = |A \sqcup B \sqcup C| &- |A \sqcup B| - |A \sqcup C| - |B \sqcup C|\\
&+ |A| + |B| + |C|.
\end{split}
\end{equation}
The similarities among these different relations might suggest that some relations are derivable from others; that information theory is somehow derivable from probability theory, that quantum mechanics is derivable from information theory, or that some of these theories can be considered to be generalizations of others \cite{Sorkin:1994, Youssef:1994, Fuchs:2002}.  Without a foundational theory explaining why any one of these different theories takes the form that it does, it is impossible to know whether one theory derives from another.

To a large degree, this paper is focused on mathematics.  Although there are immediate implications, in terms of understanding, for the sciences, specifically physics.  We seek general theories on how to quantify things.  Rather than employing the usual strategy of generalizing from specific cases, we specialize from generality \footnote{I  credit John Skilling for this insightful description.}.  Since any general theory must apply to specific cases, we may employ eliminative induction \cite{Caticha2012:entropic} by selecting simple cases that serve to rule out large classes of general theories thus severely restricting the remaining class of possible general theories.  The procedure is to repeatedly consider special cases until either the class of possible theories consist of a single theory (possibly up to isomorphism) or it is found that there is no general theory.

I consider a mathematical construct known as a lattice, and focus on the problem of consistently quantifying lattice elements by defining a function that takes lattice elements to real numbers.  Lattices exhibit various symmetries, which when considered through the application of eliminative induction constrain all attempts at quantification resulting in constraint equations, which we recognize as rules or laws.  For example, \emph{all} lattices exhibit associativity, and as a result, for any quantification there will be a constraint equation isomorphic to additivity, which is typically referred to as the \emph{sum rule}, or the inclusion-exclusion relation \cite{Rota:combinatorics, Krishnamurthy:combinatorics, Klain&Rota}.  Distributive lattices exhibit distributivity, which leads to a \emph{product rule}.

This paper builds on, and improves, past efforts \cite{Knuth:laws, Knuth:duality, Knuth:WCCI06, Knuth:lattices+probability, Knuth:measuring} by first demonstrating that associativity of the lattice join restricts quantification to be Abelian.  I then, following the lead of Cox \cite{Cox:1946} and others \cite{Smith&Erickson:associativity:1990, Garrett:1998, Jaynes:Book}, employ the functional equation known as the associativity equation to prove that quantification must exhibit properties that are isomorphic to addition.  This is, in fact, \emph{why we sum the numbers of things when we combine them} (as long as the act of combination is closed and associative).

In this case, the answer to Wigner's \cite{Wigner:1960}, Hamming's \cite{Hamming:1980} and my \cite{Knuth:FQXI2015} queries regarding the effectiveness of mathematics is that \emph{mathematics is implicitly engineered to work}.  The sum and product rules are constraint equations that enforce consistency of quantification so that the mathematics is assured to work in all situations that exhibit the requisite symmetries.  The result is a foundation of quantification that enables us to understand why many of our theories take the mathematical forms that they do.

There are some advantages to presenting these ideas in the context of order-theoretic lattices, mainly the facts that lattices are easily visualized and that many familiar problems are readily modeled as lattices.  However, there is also a serious risk in that the reader could be left with the mistaken impression that the arguments and proofs are restricted to the domain of order-theoretic lattices, and that lattices are somehow central.  Whereas the reality of the situation is that it is not that these problems can be modeled as lattices, so much that it is that these problems exhibit critical, yet common, symmetries \cite{GKS:PRA, GK:Symmetry, Knuth+Skilling:2012}.  So while lattices are the focus of this paper, it should be understood that it is really the symmetries of associativity and distributivity that are of central importance to the results herein.

In many ways this paper summarizes and brings together elements of the author's past work on quantification and measuring \cite{Knuth:laws}, the consistent quantification of lattices \cite{Knuth:duality, Knuth:lattices+probability, Knuth:probability, Knuth:me08, Knuth:measuring}, the foundation of probability theory \cite{Knuth:PhilTrans, Knuth:duality, Knuth:lattices+probability, Knuth:probability, Knuth:me08, Knuth+Skilling:2012}, a calculus for questions \cite{Knuth:PhilTrans, Knuth:me2004, Knuth:AISTATS2005, Knuth:duality, Knuth:WCCI06, Knuth:me08}, and derivation of the Feynman rules of quantum mechanics \cite{GKS:PRA, GK:Symmetry}.  Rather than following the previous approaches in which one advances quickly to the associativity equation, which results in the sum rule, this paper takes a new approach by considering quantification in more generality and demonstrating first that lattice joins result in an Abelian constraint, which is then examined in the context of the associativity equation.  The paper is organized as follows.  Section \ref{sec:lattices} provides a brief overview of lattices and their properties.  This is followed by Section \ref{sec:quantification} which introduces the concept of consistent quantification along with the more specialized concepts of valuations and co-valuations.  Section \ref{sec:symmetries} discusses important algebraic symmetries and the constraints that they place on consistent quantification.  These constraints are related in the following sections in which it is demonstrated that the resulting constraint equation on the quantification of the lattice join is abelian.  This result is then considered from the perspective of the associativity equation, which results in an additive constraint, which is implied by the fact that the constraint is abelian.  The results are then extended to the lattice product, and chaining of bi-valuations, which are then related to probability theory and other applications.

\section{Lattices and their Symmetries} \label{sec:lattices}
\black

A partially ordered set $(P,\leq)$, or \emph{poset}, is a set of elements $P$ along with a binary ordering relation, generally denoted $\leq$, which, for elements $\ea, \eb,$ and $\ec \in P$, satisfies:
\begin{align}
P1. \quad & \ea \leq \ea & \; \mbox{(Reflexivity)} \nonumber \\
P2. \quad  & \;\mbox{if}\; \ea \leq \eb \;\mbox{and}\; \eb \leq \ea \;\mbox{then}\; \ea = \eb  & \; \mbox{(Antisymmetry)} \nonumber \\
P3. \quad  & \;\mbox{if}\; \ea \leq \eb \;\mbox{and}\; \eb \leq \ec \;\mbox{then}\; \ea \leq \ec & \; \mbox{(Transitivity).} \nonumber
\end{align}
For all elements $\ea, \eb \in P$ we have that either $\eb$ includes $\ea$, denoted $\ea \leq \eb$, or $\ea$ includes $\eb$, denoted $\eb \leq \ea$, or $\ea$ and $\eb$ are incomparable, denoted $\ea \|| \eb$.  It is for this reason, that there possibly exist pairs of elements that cannot be ordered, that $(P, \leq)$ is called a \emph{partially} ordered set.

A \emph{lattice} $(L, \leq)$ is a poset where each pair of elements $\ea, \eb \in L$ has both a unique least upper bound, or supremum, called the \emph{join}, denoted $\ea \join \eb$, and a unique greatest lower bound, or infimum, called the \emph{meet}, denoted $\ea \meet \eb$.  Since the supremum and infimum both exist, the join and meet may be considered to be binary operations, $\join$ and $\meet$, that obey certain symmetries.  For example, for $\ea, \eb$ and $\ec \in L$ the join and meet operations satisfy the properties, L1 through L5 in Table \ref{tab:lattice-properties}, of idempotency, absorption, commutativity and associativity as well as the consistency relation, which relates the order-theoretic aspects of the lattice to its algebraic aspects \cite{Birkhoff:1967, Davey&Priestley}
\begin{table}[H]
\caption{A Table of five important lattice properties along with the distributive property of distributive lattices}
\begin{tabular}{l p{0.25\textwidth} r}
\midrule
\multirow{ 2}{*}{L1.} & $\ea \join \ea = \ea$ & \multirow{ 2}{*}{(Idempotency)}\\
{} & $\ea \meet \ea = \ea$ & {} \\
\midrule
\multirow{ 2}{*}{L2.} & $\ea \join (\ea \meet \eb) = \ea$ & \multirow{ 2}{*}{(Absorption)}\\
{} & $\ea \meet (\ea \join \eb) = \ea$ & {} \\
\midrule
\multirow{ 2}{*}{L3.} &$\ea \join \eb = \eb \join \ea$ & \multirow{ 2}{*}{(Commutativity)}\\
{} & $\ea \meet \eb = \eb \meet \ea$ & {} \\
\midrule
\multirow{ 2}{*}{L4.} & $\ea \join (\eb \join \ec) = (\ea \join \eb) \join \ec$ & \multirow{ 2}{*}{(Associativity)}\\
{} & $\ea \meet (\eb \meet \ec) = (\ea \meet \eb) \meet \ec$. & {}\\
\midrule
L5. & $\ea \leq \eb \Leftrightarrow
\begin{matrix}
\ea \meet \eb = \ea \\ \ea \join \eb = \eb
\end{matrix}$ & (Consistency)\\
\midrule
\midrule
D1. & $\ea \join (\eb \meet \ec) = (\ea \join \eb) \meet (\ea \join \ec)$ & \multirow{ 2}{*}{(Distributivity)}\\
D2. & $\ea \meet (\eb \join \ec) = (\ea \meet \eb) \join (\ea \meet \ec)$ & {}\\
\midrule
\end{tabular}\label{tab:lattice-properties}
\end{table}
\noindent
Distributive lattices exhibit the additional properties D1 and D2 where the join distributes over the meet, and vice versa.  The dual relations are related by reversing the ordering relation, or equivalently, by interchanging join and meet.  Since the join and meet operations obey algebraic relations, \emph{every lattice is an algebra}.

If the poset is such that each pair of elements has a supremum, but not necessarily an infimum, then it is called a \emph{join-semilattice}.  The \emph{meet-semilattice} is defined dually.

In this work we focus on locally finite lattices \red and join-semilattices \black in which every closed interval $[\ea,\eb] = \{ \ex : \ea \leq \ex \leq \eb \}$ is finite. \red Continuous lattices need not be considered since one cannot measure infinitesimal differences in practice.  In effect, each element represents an equivalence class of objects selected for a desired application.  For example, one element may represent apples and another element may represent oranges.  An infinite number of elements, or equivalence classes, need not be considered.  Practically, one would have neither the time nor space to identify, store or address an infinite number of equivalence classes. Therefore, a finite, albeit possibly extremely large, number of equivalence classes will always suffice allowing anyone to describe a set of objects to within requisite precision.  For this reason, it suffices to focus on locally finite lattices and join-semilattices. \black

Here we consider maps called \emph{quantifications} that take elements of lattices, join-semilattices, or meet-semilattices to numbers that are totally ordered, such as integers or reals.  This can be motivated by the desire to rank a partially ordered set by mapping it to a total order.  To do this one can employ a special class of quantifications called a \emph{valuation} for which given elements $\ea, \eb \in L$, the valuation $v$ takes $\ea$ and $\eb$ to numbers $v(\ea)$ and $v(\eb)$ such that $\ea \leq \eb$ implies that $v(\ea) \leq v(\eb)$.\footnote{Note that the symbol `$\leq$' is overloaded so that when comparing lattice or poset elements, as in $\ea \leq \eb$, it represents the binary ordering relation, and when comparing quantifications or valuations (real numbers), as in $v(\ea) \leq v(\eb)$, it represents the usual less-than-or-equal-to comparator.}  \red Assignment of a valuation to the lattice elements by the function $v$ serves to rank the elements of the lattice via an order-preserving map that is referred to as \textit{\emph{fidelity}} \cite{Knuth+Skilling:2012}.  Relaxing the fidelity requirement results in a more general quantification that can have both positive and negative values, which is often referred to as a signed measure. \black

One can also choose to assign a dual ranking using a \emph{co-valuation} where $\ea \leq \eb$ implies that $v(\ea) \geq v(\eb)$.  In other applications, one may find it useful to quantify the lattice using neither a valuation nor a co-valuation.  Figure 1 illustrates four examples of a quantification of a chain (totally ordered set).

\begin{figure}
\centering
\makebox{\includegraphics[width=\columnwidth]{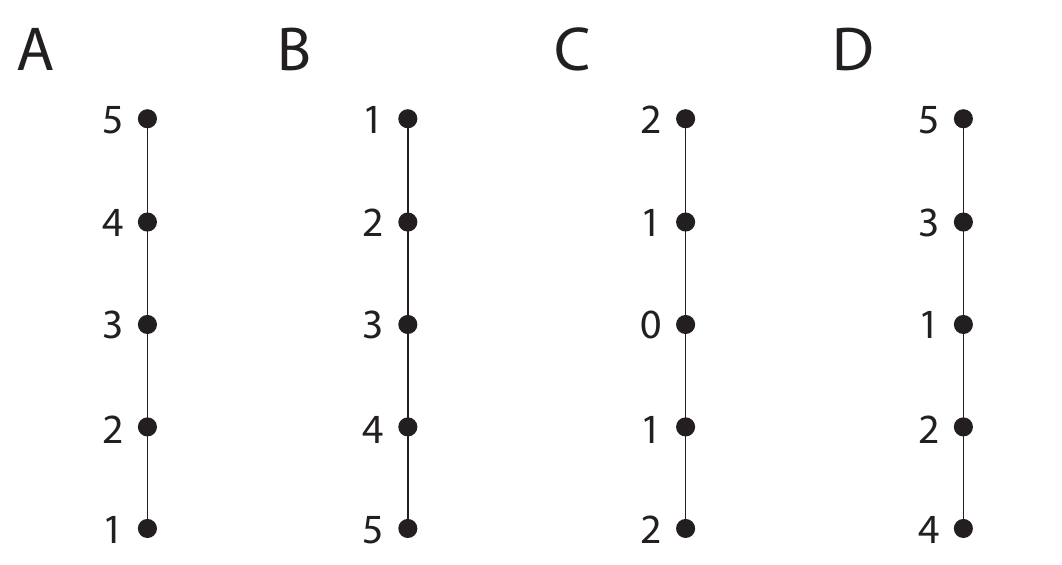}}
\caption{This figure illustrates several quantifications of a chain of five elements: A. A valuation that relies on the natural numbers from one to five. B. A co-valuation that relies on the natural numbers from one to five.  C. A quantification that takes advantage of some aspect of the total order, yet is neither a valuation nor a co-valuation.  The utility of such a quantification is readily apparent to anyone who has ever considered their distance from a destination, such as a rest stop along a highway.  D. An inconsistent quantification that does not appear to encode any aspect of the total order.} \label{fig:posets}
\end{figure}

\section{Consistent Quantification of Lattices} \label{sec:quantification}
\begin{figure}
\centering
\makebox{\includegraphics[width=0.75\columnwidth]{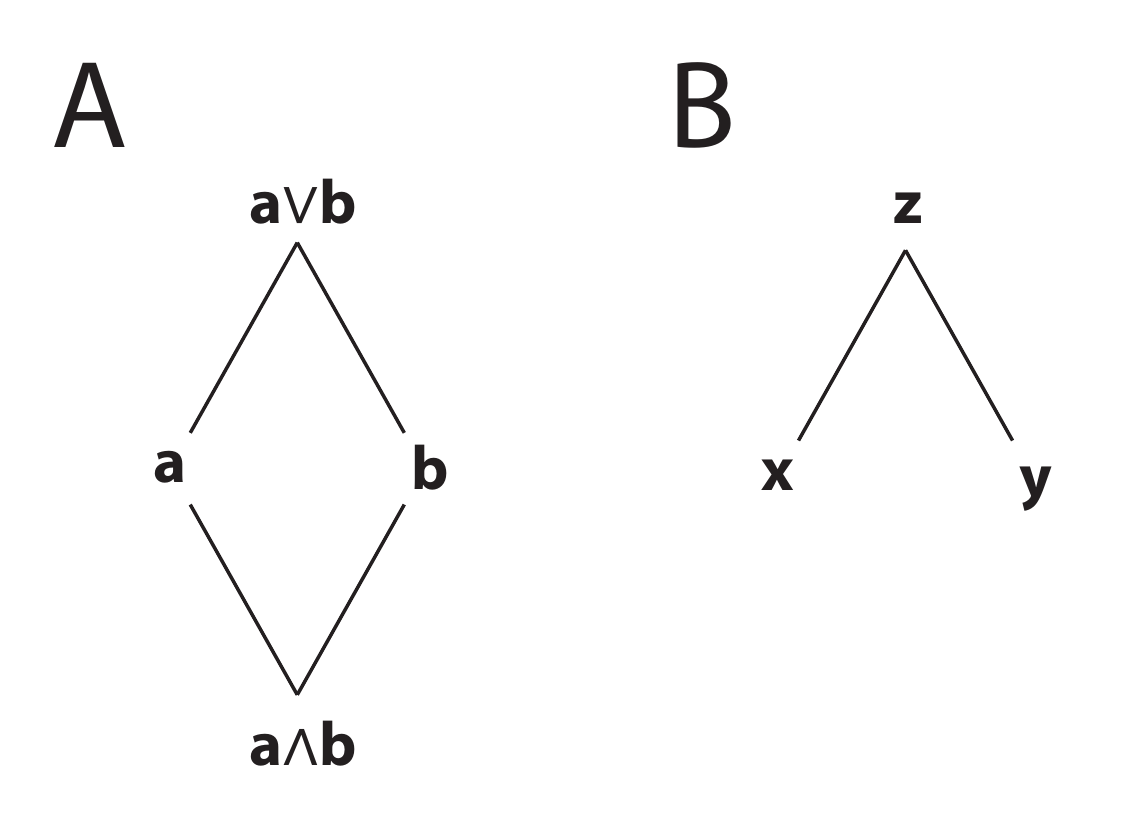}}
\caption{This figure illustrates the two main structures considered in the derivation of consistent quantification.  A. This is the lattice with elements $\ea$ and $\eb$ along with their join $\ea \vee \eb$ and meet $\ea \wedge \eb$.  B.  This is the join semi-lattice consisting of elements $\ex$ and $\ey$ and their join $\ez = \ex \vee \ey$.  The elements $\ex$ and $\ey$ can be considered mutually exclusive since their meet is null.} \label{fig:join+oplus}
\end{figure}

For a quantification to encode some aspect of the lattice structure, which we refer to as \emph{consistent quantification}, one would expect that since lattice elements $\ea$ and $\eb \in L$ are related to elements $\ea \wedge \eb$ and $\ea \vee \eb \in L$ (Figure \ref{fig:join+oplus}A), then there ought to be a functional relationship among the quantities assigned to the set of elements $\ea$, $\eb$, $\ea \vee \eb$ and $\ea \wedge \eb$.  We postulate a function $F$ that relates the quantification $q(\ea \vee \eb)$ assigned to the element $\ea \vee \eb$ to the quantifications $q(\ea)$, $q(\eb)$, and $q(\ea \wedge \eb)$ assigned to elements $\ea$, $\eb$, and $\ea \wedge \eb$, respectively by
\begin{equation} \label{eq:F_1}
q(\ea \vee \eb) = F( q(\ea), q(\eb), q(\ea \wedge \eb) ),
\end{equation}
which by defining $a = q(\ea)$, $b = q(\eb)$, $c = q(\ea \wedge \eb)$, $d = q(\ea \vee \eb)$,
can be compactly written as
\begin{equation} \label{eq:F_2}
d = F( a, b, c ).
\end{equation}
Our aim is to identify which set of functions $F$ satisfy the relevant constraints.

We also consider a join semi-lattice (Figure \ref{fig:join+oplus}B) with three elements $\ex$, $\ey$ and $\ez$, such that $\ez = \ex \join \ey$, and $\ex$ and $\ey$ are disjoint such that the element $\ex \meet \ey$ is null, and thus has been omitted.\footnote{It is not unusual in order theory for the null (bottom) element of a lattice to be omitted.}   The concept of consistent quantification requires that the quantification should carry some information about the order-theoretic relation among the elements.  In this specific case, since we have that the element $\ez = \ex \join \ey$, the quantity $z = q(\ez)$ assigned to the element $\ez$ must be some function of the quantities $x = q(\ex)$ and $y = q(\ey)$ assigned to elements $\ex$ and $\ey$, respectively.  We write this relationship as
\begin{equation} \label{eq:oplus}
z = x \oplus y,
\end{equation}
in which $\oplus$ is a real-valued binary operator to be determined.

\red
With these concepts in mind, we can formally define a consistent quantification of a lattice and a join semi-lattice,
\begin{definition}[Consistent Quantification]
A consistent quantification of a lattice, or a join semi-lattice, is a function $q$ that takes every element $\ex \in L$ to a real number $q(\ex) \in \mathbb{R}$, such that for all $\ea, \eb, \ea \join \eb, \ea \meet \eb \in L$ there exists a real-valued function $F$ with which $q(\ea \vee \eb) = F( q(\ea), q(\eb), q(\ea \wedge \eb) )$, or in the case for which there does not exist an element $\ea \meet \eb$ there exists a real-valued binary operator $\oplus$ for which $q(\ea \join \eb) = q(\ea) \oplus q(\eb)$.
\end{definition}
We will later require the quantification assignments made by function $F$ to agree with those made by the operator $\oplus$ in the case for which the bottom element of a lattice is null so that it can be optionally neglected resulting in a join semi-lattice.
\black

In the following sections, we will rely on symmetries and special cases to restrict the possible forms of the function $F$, the related operator $\oplus$, and their relationship to one another.  The special cases will rely on the fact that the definition of consistent quantification removes one degree of freedom thus enabling one to freely assign three of the four quantifications $q(\ea), q(\eb), q(\ea \join \eb)$ and $q(\ea \meet \eb)$ and two of the three quantifications $q(\ex), q(\ey)$ and $q(\ex \join \ey)$ in the case where $\ex \meet \ey$ does not exist.

\section{Symmetries} \label{sec:symmetries}
General rules must hold for special cases.  Here we proceed by using eliminative induction \cite{Caticha2012:entropic}, which consists of identifying simple special cases that rule out, or eliminate, possible forms for $F$ and $\oplus$.  We begin by considering some basic symmetries.

\subsection{Commutativity}
In general, we have commutativity of the join and meet (L3),
which results in
\begin{equation} \label{eq:sym-commutativity}
d = F(a,b,c) = F(b,a,c),
\end{equation}
since $c = q(\ea \wedge \eb) = q(\eb \wedge \ea)$ and $d = q(\ea \vee \eb) = q(\eb \vee \ea)$.

In addition, commutativity of the join, $\ex \vee \ey = \ey \vee \ex$, enables us to write (\ref{eq:oplus}) as
\begin{equation}
z = x \oplus y = y \oplus x,
\end{equation}
so that the real-valued binary operator $\oplus$ must also be commutative.

\subsection{Associativity}
In addition to being commutative, the join and meet operators are also associative (L4), so that for disjoint elements $\ew$, $\ex$, and $\ey$, the relation
\begin{equation}
\ew \vee (\ex \vee \ey) \equiv (\ew \vee \ex) \vee \ey
\end{equation}
implies that the quantifications satisfy
\begin{equation} \label{eq:associativity}
w \oplus (x \oplus y) = (w \oplus x) \oplus y,
\end{equation}
so that the operator $\oplus$ is also associative.  The relation (\ref{eq:associativity}) is a functional equation for the operator $\oplus$ known as the \emph{associativity equation}, which will be discussed in Section \ref{sec:associativity-equation}.

\begin{figure}
\centering
\makebox{\includegraphics[width=0.4\columnwidth]{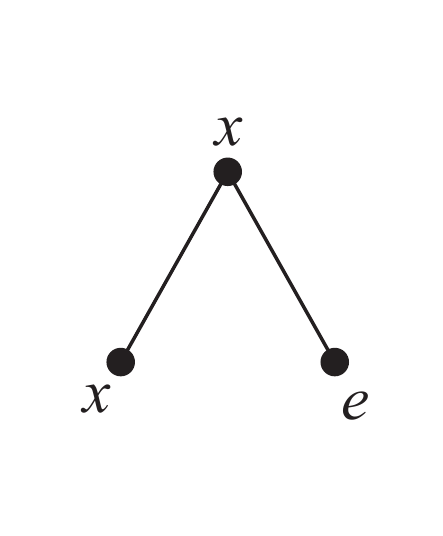}}
\caption{This figure illustrates the simple join-semilattice in Figure \ref{fig:join+oplus}B where the elements are quantified so that $q(\ex) = x$, $q(\ez) = x$, and $q(\ey) = e$.  This sets up the situation where the real number $e$ is the identity element for the operator $\oplus$.} \label{fig:identity}
\end{figure}

Naturally, associativity also constrains the function $F$ by requiring that
\begin{equation}
F(F(a,b,c),f,g) = F(a,F(b,f,g),c),
\end{equation}
where we have used $(\ea \join \eb) \join \ef = \ea \join (\eb \join \ef)$ with $\ea \meet \eb = \ec$, and $\eb \meet \ef = \eg$. This can be made more symmetric by using commutativity and relabeling
\begin{equation}
F(F(a,b,c),f,g) = F(F(a,f,g),b,c).
\end{equation}
However, it will be more profitable to proceed by first relating the function $F$ to the operator $\oplus$.

\section{Relating $\oplus$ to $F$}
We now focus on a special case of the join-semilattice illustrated in Figure \ref{fig:join+oplus}B where the elements are chosen to be quantified so that $q(\ex) = x$, $q(\ez) = x$, and $q(\ey) = e$ where $x$ and $e$ are real numbers as illustrated in Figure \ref{fig:identity}.  Such a quantification must satisfy
\begin{equation}
x \oplus e = x
\end{equation}
for all values of $x$ so that for this quantification to be a consistent quantification, the real number $e$ must be the identity element for the operator $\oplus$.

\begin{figure}
\centering
\makebox{\includegraphics[width=0.4\columnwidth]{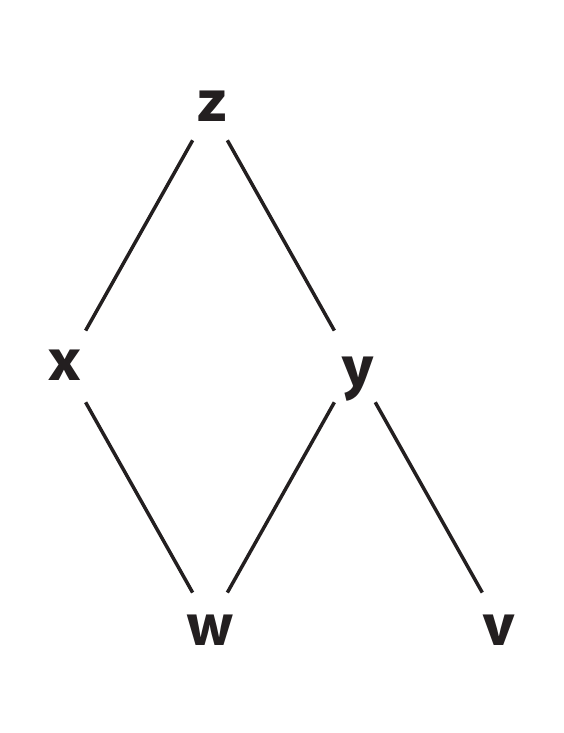}}
\caption{This figure illustrates a lattice structure appended to a join-semilattice so that the function $F$ can be related to the operator $\oplus$ by $x \oplus v = F(x, w \oplus v, w)$.} \label{fig:F-oplus-relation}
\end{figure}

The next task is to relate the function $F$ to the operator $\oplus$ by building on the join semi-lattice forming the structure illustrated in Figure \ref{fig:F-oplus-relation}.  It is easily verified that
\begin{equation}
z = x \oplus v
\end{equation}
and
\begin{equation}
z = F(x,y,w)
\end{equation}
so that
\begin{equation} \label{eq:equating-F-with-oplus}
x \oplus v = F(x,y,w).
\end{equation}
Moreover, it is also true that
\begin{equation}
y = w \oplus v
\end{equation}
so that (\ref{eq:equating-F-with-oplus}) becomes
\begin{equation} \label{eq:oplus-to-F}
x \oplus v = F(x, w \oplus v, w)
\end{equation}
for all values of $v$, $w$, $x$ and $y$.

\begin{figure}
\centering
\makebox{\includegraphics[width=1\columnwidth]{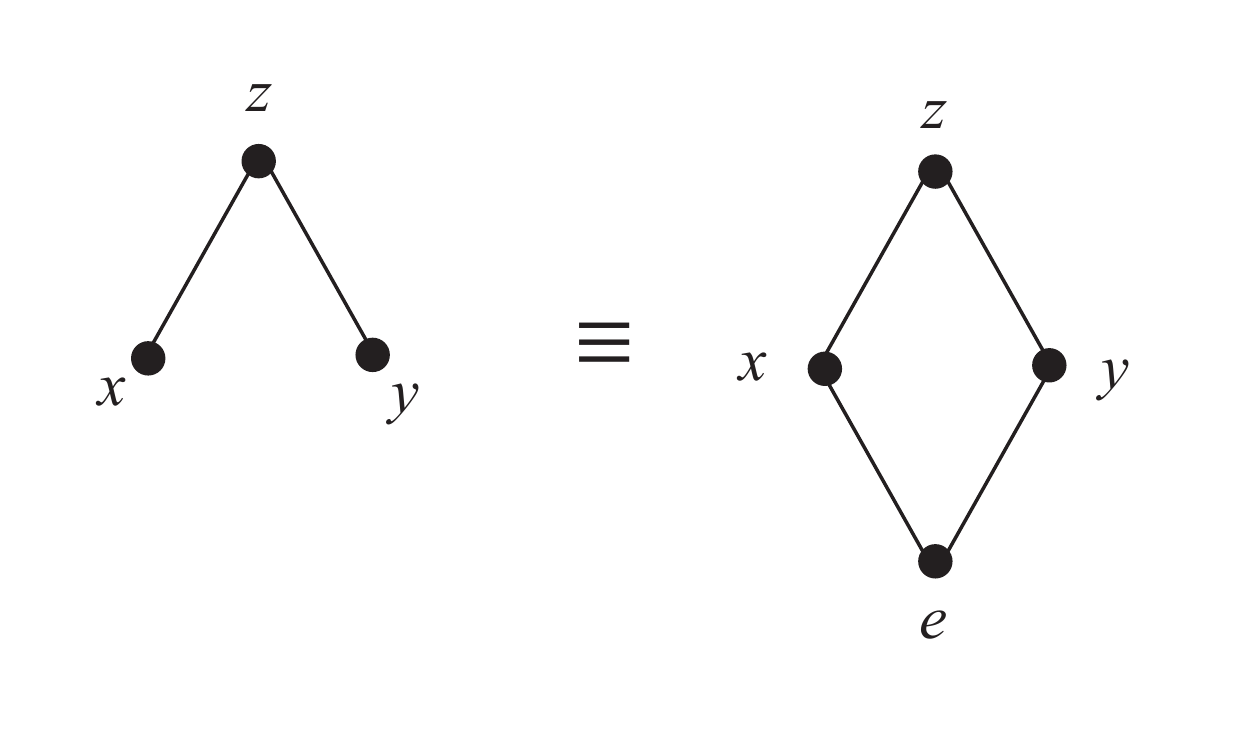}}
\caption{This figure illustrates that quantifying the bottom element with the identity $e$ is equivalent to neglecting the bottom since $x \oplus y = F(x, y, e)$.} \label{fig:bottom-equivalence}
\end{figure}

If we now consider the special case where the element $\ew$ is quantified by the identity, $w = e$, we then have that $w \oplus v = e \oplus v = v$ and (\ref{eq:oplus-to-F}) becomes
\begin{equation}
x \oplus v = F(x, v, e).
\end{equation} \red
Thus assigning the identity $e$ of $\oplus$ to the bottom (null) element of a lattice is equivalent to neglecting the bottom element and representing the structure as a join semi-lattice (see Figure \ref{fig:bottom-equivalence}). \black

\section{$\oplus$-Inverse} \red
In this section, we demonstrate that the operator $\oplus$ must have an inverse operation.  This does not imply that every join semi-lattice, or lattice, must have elements that are quantified by both numbers and their inverses (under $\oplus$).  Whether inverse elements are necessary for quantification in a given application is dependent both on the specific lattice structure and the assignments made to the join-irreducible elements.  What is important is that the rules $\oplus$ and $F$ for relating quantifications can accommodate inverses.
\black

\begin{figure}
\centering
\makebox{\includegraphics[width=0.4\columnwidth]{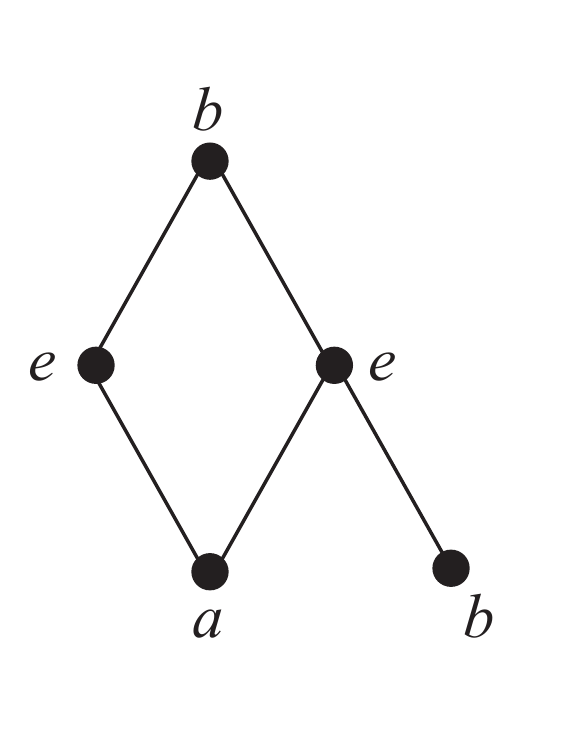}}
\caption{This figure illustrates the same structure in Figure \ref{fig:F-oplus-relation}, but quantified to illustrate that the $\oplus$-inverse of $a$ is given by $F(e,e,a)$.} \label{fig:oplus-inverse}
\end{figure}

Consider the structure in Figure \ref{fig:F-oplus-relation} with a particular quantification, illustrated in Figure \ref{fig:oplus-inverse}, such that $x = y = e$, $z = b$ and $w = a$.  It is then clear that since $v \oplus x = z$, we have that $v \oplus e = b$, which implies that $v = b$.  

Since $\ew \vee \ev = \ey$, we have that
\begin{equation}
a \oplus b = e.
\end{equation}
This constrains the relationship between the quantifications $a$ and $b$, so that for the given quantification assignments to be a consistent quantification, it must be that
the operator $\oplus$ must support an inverse operation, such that $b$ is the $\oplus$-inverse of $a$, which we will write as $a^{-1}$:
\begin{equation}
b = a^{-1}.
\end{equation}
This enables us to define the inverse operator $\ominus$ for which
\begin{equation}
e \ominus a = a^{-1}.
\end{equation}

However, from Figure \ref{fig:oplus-inverse}, we also have that
\begin{equation}
b = F(e,e,a)
\end{equation}
so that
\begin{equation}
a^{-1} = e \ominus a = F(e,e,a).
\end{equation}


\section{Additivity}
We now again consider the structure illustrated in Figure \ref{fig:F-oplus-relation}.  We have that
\begin{align}
z &= x \oplus v\\
y &= w \oplus v,
\end{align}
which implies that
\begin{equation}
y \ominus w = v
\end{equation}
so that
\begin{equation}
z = x \oplus (y \ominus w).
\end{equation}
This allows us to write
\begin{equation}
F(x,y,w) = x \oplus (y \ominus w),
\end{equation}
and since $\oplus$ is associative, we have
\begin{equation}
F(x,y,w) = x \oplus y \ominus w,
\end{equation}
which is a manifestation of the inclusion-exclusion principle of combinatorics \cite{Rota:combinatorics, Krishnamurthy:combinatorics, Klain&Rota}.
We have demonstrated that the operator $\oplus$ is Abelian, which means that it forms a group such that it has an identity $e$, every element has an inverse, and $\oplus$ is associative and commutative.
One possible solution for $\oplus$ is addition, so that we can write the \emph{Sum Rule} as
\begin{equation}
F(x,y,w) = x + y - w,
\end{equation}
which is not surprising since Abelian groups represent generalized addition.

This analysis served to establish the fact that $\oplus$ is Abelian without resorting to functional equations, which can be rather obscure.  In the next section, we will discuss the solution to the associativity equation in (\ref{eq:associativity}) and show that without loss of generality, we can always choose to quantify the lattice so that the operator $\oplus$ is represented by addition.

It has been suggested that one possible solution, consistent with the bespoke symmetries, is the $\max$ function \cite{Teran:2015}
$$
x \oplus y = \max{(x,y)},
$$
which clearly cannot be regraded to standard addition.  However, this suggestion fails, not only in the cases where one aims to rank elements, but also in general because there is no possibility of an identity element and inverse elements for which $a \oplus a^{-1} = \max{(a,a^{-1})} = e$ could be satisfied.

\section{The Associativity Equation}  \label{sec:associativity-equation}
In this section, we consider the associativity equation (\ref{eq:associativity}), and present (paraphrase) a theorem
from Acz\'{e}l \cite{Aczel:FunctEqns}:
\begin{theorem}[Associativity Theorem]
If $x \oplus y \in (a,b)$ is continuous with respect to $x \in (a,b)$ and $y \in (a,b)$, and satisfies
$$
x \oplus (y \oplus z) = (x \oplus y) \oplus z
$$
for each value $x, y, z \in (a,b)$, and if there exist in $(a,b)$ real numbers $e$ and $x^{-1}$ such that
$$
e \oplus x = x
$$
and
$$
x^{-1} \oplus x = e
$$
hold, then and only then does there exist a continuous and strictly monotonic function $f(x)$ defined in $(-\infty, \infty)$, with range $(a,b)$ and with inverse $f^{-1}(x)$, such that
\begin{equation}
x \oplus y = f \left( f^{-1}(x) + f^{-1}(y) \right).
\end{equation}
\end{theorem}

\red
There appears to be no unique minimal set of conditions that lead to additivity.
For example, similar theorems have been presented and proven by Azc\'{e}l \cite{Aczel:associativity:2004}, and Craigen and Pales \cite{Craigen+Pales:1989}, and Knuth and Skilling \cite{Knuth+Skilling:2012} in which, given disjoint $\ex$, $\ey$, and $\ez$, cancellativity
\begin{equation}
\ex \begin{Bmatrix} < \\ = \\ > \end{Bmatrix} \ey  \qquad \mbox{implies} \qquad \ex \oplus \ez \begin{Bmatrix} < \\ = \\ > \end{Bmatrix} \ey \oplus \ez
\end{equation}
which formalizes a concept of ranking,
was postulated in lieu of postulating an identity and inverse in $(a,b)$. \black

For present purposes, the main result is that the function relating the quantities $x$ and $y$ assigned to two disjoint elements $\ex$ and $\ey$ to the quantity $x \oplus y$ assigned to their join $\ex \vee \ey$ can be expressed as an invertible transform of ordinary addition
\begin{equation} \label{eq:oplus-solution}
x \oplus y = f \left( f^{-1}(x) + f^{-1}(y) \right).
\end{equation}
where $f$ is an \emph{arbitrary} invertible function.  This can be viewed as a constraint equation, which ensures that associativity is satisfied by the assigned valuations.  Given the linearity of this associativity constraint (\ref{eq:oplus-solution}), the only remaining freedom is that of rescaling.

This means that given any consistent quantification with a definition of $\oplus$, one can rescale, or \emph{regraduate}, the quantification by mapping the quantity $q$ to a new quantity $g(q)=f^{-1}(q)$ so that the addition holds
\begin{equation} \label{eq:regraduated-sum-rule}
g(a \oplus b) = g(a) + g(b).
\end{equation}
By using the quantifications $g(a)$, $g(b)$, and $g(a \oplus b)$ instead of $a$, $b$, and $a \oplus b$, we can adopt $+$ instead of another Abelian operator $\oplus$.  Thus for
$\ez = \ex \join \ey$
in which $\ex$ and $\ey$ are disjoint
one can always assign quantities $x$, $y$, and $z$, such that
\begin{equation} \label{eq:oplus-sum}
z = x + y,
\end{equation}
which is the sum rule in the case of disjoint elements.

More generally, for $\ez = \ex \join \ey$ and $\ew = \ex \meet \ey$ we can write the sum rule as
\begin{equation} \label{eq:join-sum}
z = x + y - w.
\end{equation} \red
Since all lattices have a join operation that is commutative and associative, the sum rule holds for all lattices. However, it should be noted that while the sum rule holds for all lattices, it cannot be assured that the resulting quantification will be a valuation.  That is, it is not generally true that for elements $\ex \leq \ey$ we will have $q(\ex) \leq q(\ey)$. An example of this is the co-information lattice in information theory \cite{Bell:2003} and relevance among questions \cite{vanErp+etal:2017} for which some quantities are negative.\black

\section{Lattice Products}
Lattices can be combined using the Cartesian product, or \emph{lattice product}.  That is, given two lattices $A$ and $B$, one can define a partial order over the lattice product $A \times B$.  This is accomplished by considering elements $(\ea_1, \eb_1)$ and $(\ea_2, \eb_2) \in A \times B$ and defining $(\ea_1,\eb_1) \leq (\ea_2, \eb_2)$ iff $\ea_1 \leq \ea_2$ and $\eb_1 \leq \eb_2$.

Given a quantification $q$ of lattices $A$ and $B$, in which the element $\ea \in A$ is quantified by $a \equiv q(\ea)$ and the element $\eb \in B$ is quantified by $b \equiv q(\eb)$, we consider the quantity that should be assigned to the element $(\ea,\eb) \in A \times B$.  Consistency requires that the number assigned to the element $(\ea,\eb)$ must be a function of the numbers assigned to the elements $\ea$ and $\eb$:
\begin{equation}
q((\ea,\eb)) = q(\ea) \otimes q(\eb)
\end{equation}
in which the real-valued binary operator $\otimes$ is to be determined.

The lattice product is associative, so that $(A \times B) \times C = A \times (B \times C)$.  As a result, we have that
\begin{equation}
\big( (\ea, \eb), \ec \big) = \big( \ea, (\eb, \ec) \big),
\end{equation}
which implies that the operator $\otimes$ is associative
\begin{equation}
(a \otimes b) \otimes c = a \otimes (b \otimes c).
\end{equation}

The lattice product obeys cancellativity since given $\ea$ and $\ec \in A$ where $\ea \leq \ec$ and given $\eb \in B$, it is true that $(\ea,\eb) \leq (\ec,\eb)$ since $\ea \leq \ec$ and $\eb \leq \eb$.
By the theorems in \cite{Craigen+Pales:1989} and \cite{Knuth+Skilling:2012}, we have that the operator $\otimes$ is an invertible transform of addition
\begin{equation} \label{eq:otimes}
a \otimes b = h^{-1}(h(a) + h(b)).
\end{equation}

The lattice product is distributive over the lattice join.  That is,
given disjoint $\ea_1$ and $\ea_2 \in A$, and $\eb \in B$, we then have that
\begin{equation} \label{eq:product-distributive-over-join}
(\ea_1 \join \ea_2, \eb) = (\ea_1, \eb) \join (\ea_2, \eb).
\end{equation}
From (\ref{eq:otimes}) we have that
\begin{align}
q\big( (\ea_1 , \eb) \big) &= h^{-1}\Big(h\Big(q\big(\ea_1\big)\Big) + h\Big(q\big(\eb\big)\Big)\Big) \\
q\big( (\ea_2 , \eb) \big) &= h^{-1}\Big(h\Big(q\big(\ea_2\big)\Big) + h\Big(q\big(\eb\big)\Big)\Big) \\
q\big( (\ea_1 \join \ea_2, \eb) \big) &= h^{-1}\Big(h\Big(q\big(\ea_1 \join \ea_2\big)\Big) + h\Big(q\big(\eb\big)\Big)\Big).
\end{align}
Now by writing
\begin{align}
x &= h\Big( q\big(\ea_1\big) \Big) \\
y &= h\Big( q\big(\ea_2\big) \Big) \\
z &= h\Big( q\big(\eb\big) \Big) \\
k(x,y) &= h\Big( q\big(\ea_1 \join \ea_2\big) \Big),
\end{align}
and letting $H(x) = h^{-1}(x)$ we have that (\ref{eq:product-distributive-over-join}) implies that
\begin{equation} \label{eq:product-equation}
H( k(x,y) + z ) = H(x+z) + H(y+z),
\end{equation}
which is a functional equation known as the \emph{product equation} as it encodes the fact that the lattice product is distributive over the join.
The solution of the product equation (\ref{eq:product-equation}) is that \cite{Knuth+Skilling:2012}
\begin{equation}
h(x) = \log(x)
\end{equation}
so that the operator $\otimes$ (\ref{eq:otimes}) is multiplication with
\begin{equation}
q((\ea,\eb)) = C q(\ea) q(\eb),
\end{equation}
in which $C$ is an arbitrary positive constant, which amounts to a choice of units.  The constant $C$ can be set to unity without loss of generality resulting in the
the \emph{direct product rule}
\begin{equation} \label{eq:direct-product-rule}
q((\ea,\eb)) = q(\ea) q(\eb).
\end{equation}
\red This direct product rule is what is used when one analyzes two problems jointly and assigns, for example, a joint probability to the product space based on the product of two separate probability distributions for each factor space. \black

The fact that the operator $\otimes$ can only be multiplication could have been reasoned by considering that when addition was selected for $\oplus$ in the case of the product of two disjoint lattice elements, there remained only one degree of freedom in which the quantifications could be rescaled.  The fact that quantifications can only be rescaled implies that the only possible operations consistent with summation under the lattice product are multiplicative.

\section{Bi-Quantifications}
It is also interesting to consider another form of quantification called a \emph{bi-quantification}, which is a function $b$ that takes an ordered pair of elements to a real number so that $b : (\ex, \et) \rightarrow b(\ex, \et) \in \mathbb{R}$.  The second element of the pair is referred to as the context.

It is useful to conceive of bi-quantifications as quantifying the relationship between two elements. One can think of these elements $\ex$ and $\et$ as defining a directed interval $[\ex, \et]$, and the bi-quantification as quantifying that interval.  We may then also write $b([\ex, \et]) \equiv b(\ex, \et)$.

\subsection{Bi-Quantifications under Join}

By considering bi-quantifications $b([\ex, \et])$ where the context $\et$ is kept constant, we are left with a quantification with one degree of freedom $q_{\et}(\ex) = b([\ex,\et])$ that takes the element $\ex$ to a real number.  We have from (\ref{eq:oplus-sum}) that for disjoint $\ex$ and $\ey$
\begin{equation}
\ez = \ex \join \ey  \qquad \rightarrow \qquad q_{\et}(\ez) = q_{\et}(\ex) + q_{\et}(\ey),
\end{equation}
which implies that
\begin{equation}
b([\ez, \et]) = b([\ex, \et]) + b([\ey, \et]).
\end{equation}
Similarly for general $\ex$ and $\ey$ where $\ez = \ex \join \ey$ and $\ew = \ex \meet \ey$, from (\ref{eq:join-sum}) we can write
\begin{equation}
q_{\et}(\ez) = q_{\et}(\ex) + q_{\et}(\ey) - q_{\et}(\ew),
\end{equation}
which implies that
\begin{equation}
b([\ex \join \ey, \et]) = b([\ex, \et]) + b([\ey, \et]) - b([\ex \meet \ey, \et]).
\end{equation}
Thus bi-quantifications also obey the sum rule under the join of the first element.

\subsection{Bi-Quantifications: Chaining Context}
We now consider relating bi-quantifications that have different contexts.  If we think of the pair of elements as defining a directed interval, we can consider the bi-quantification that one would assign to the concatenation, or \emph{chaining}, of two directed intervals that share, at most, a common endpoint.  For example, consider an interval formed from the chaining of two intervals $[\ex, \ey]$ and $[\ey, \ez]$:
\begin{equation}
[\ex, \ez] = \big[[\ex, \ey],[\ey, \ez]\big],
\end{equation}
so that the second element, or context, of one interval is the first element of the second interval.
Consistent quantification requires that the bi-quantification assigned to the interval $[\ex, \ez]$ must be some function of the bi-quantifications assigned to each of the two intervals $[\ex, \ey]$ and $[\ey, \ez]$, which we will write with the real-valued binary operator $\odot$
\begin{equation}
b([\ex, \ez]) = b([\ex, \ey]) \odot b([\ey, \ez]),
\end{equation}
where the functional form of the operator $\odot$ is to be determined.

Clearly, since chaining intervals is associative,
\begin{align}
[\ex, \et] &= \Big[\big[[\ex, \ey],[\ey, \ez]\big],[\ez, \et]\Big] \nonumber \\
&=  \Big[[\ex, \ey],\big[[\ey, \ez],[\ez, \et]\big]\Big]
\end{align}
it must be that the function $\odot$ is associative
\begin{align}
b([\ex, \et]) &= \Big( b([\ex, \ey]) \odot b([\ey, \ez]) \Big) \odot b([\ez, \et]) \nonumber \\
&=  b([\ex, \ey]) \odot \Big( b([\ey, \ez]) \odot b([\ez, \et]) \Big).
\end{align}

The function $\odot$ must have an identity element since
\begin{equation}
[\ex, \et] = \big[[\ex, \ex],[\ex, \et]\big]
\end{equation}
and
\begin{equation}
b([\ex, \et]) = b([\ex, \ex]) \odot b([\ex, \et])
\end{equation}
so that the $\odot$-identity is given by $e_{\odot} = b([\ex, \ex])$ for all $\ex$.
While each interval does have an inverse under concatenation
\begin{equation}
[\ex, \ex] = \big[[\ex, \ey],[\ey, \ex]\big],
\end{equation}
in which $[\ey, \ex]$ is the inverse of $[\ex, \ey]$,
the intervals $[\ex, \ey]$ and $[\ey, \ex]$ share more elements than a single endpoint.  As a result, chaining intervals does not support the inverse condition.

However, concatenation does obey cancellativity, since for $\ex < \ey < \ez < \et$, we have that
$$
[\ey, \ez] < [\ex, \ez],
$$
where proper inclusion $<$ here represents subset inclusion $\subset$, which implies that
$$
\big[[\ey, \ez],[\ez, \et]\big] < \big[[\ex, \ez],[\ez, \et]\big],
$$
since
$$
[\ey, \et] < [\ex, \et].
$$
Since cancellativity holds, by \cite{Craigen+Pales:1989} and \cite{Knuth+Skilling:2012} we have that $\odot$ is additive
\begin{equation}
b([\ex, \ey]) \odot b([\ey, \et]) = g^{-1}\left( g(b([\ex, \ey])) + g(b([\ey, \et])) \right)
\end{equation}
where $g$ is an arbitrary invertible function.

Chaining is distributive, since
\begin{equation}
\big[[\ex \join \ey,\ez],[\ez,\et]\big] = \big[[\ex,\ez],[\ez,\et]\big] \join  \big[[\ey,\ez],[\ez,\et]\big],
\end{equation}
which has bi-quantification assignments
\begin{multline} \label{eq:distributive}
\big( b([\ex,\ez]) + b([\ey,\ez] \big) \odot b([\ez,\et]) \\= \big(b([\ex,\ez]) \odot b([\ez,\et])\big) + \big(b([\ey,\ez]) \odot b([\ez,\et])\big).
\end{multline}
Despite the fact that $\odot$ must be an invertible transform of addition, addition does not satisfy the above relation.
By selecting addition for the operator $\oplus$, one still has the freedom to rescale the quantification.  As a result, just as in the case of the lattice product, the only possible functional form of the operator $\odot$ is that of multiplication, which is an invertible transform of addition, as expected.

The result is that under chaining
$$
[\ex, \ez] = \big[[\ex, \ey],[\ey, \ez]\big]
$$
the bi-quantification of the resulting interval is found by taking the product of the two intervals forming the chain
\begin{equation} \label{eq:pre-chain-rule}
b(\big[[\ex, \ey],[\ey, \ez]\big]) = C b([\ex,\ey])b([\ey,\ez]),
\end{equation}
in which $C$ is an arbitrary positive constant.  Without loss of generality the overall scale of the quantification can be set by setting $C$ equal to unity so that
\begin{equation} \label{eq:chain-rule}
b(\big[[\ex, \ey],[\ey, \ez]\big]) = b([\ex,\ey])b([\ey,\ez]),
\end{equation}
which is the \emph{chain rule}.  The equation (\ref{eq:distributive}) above becomes
\begin{multline}
\big( b([\ex,\ez]) + b([\ey,\ez] \big) b([\ez,\et]) \\= \big(b([\ex,\ez]) b([\ez,\et])\big) + \big(b([\ey,\ez]) b([\ez,\et])\big).
\end{multline}

\begin{figure}
\centering
\makebox{\includegraphics[width=0.5\columnwidth]{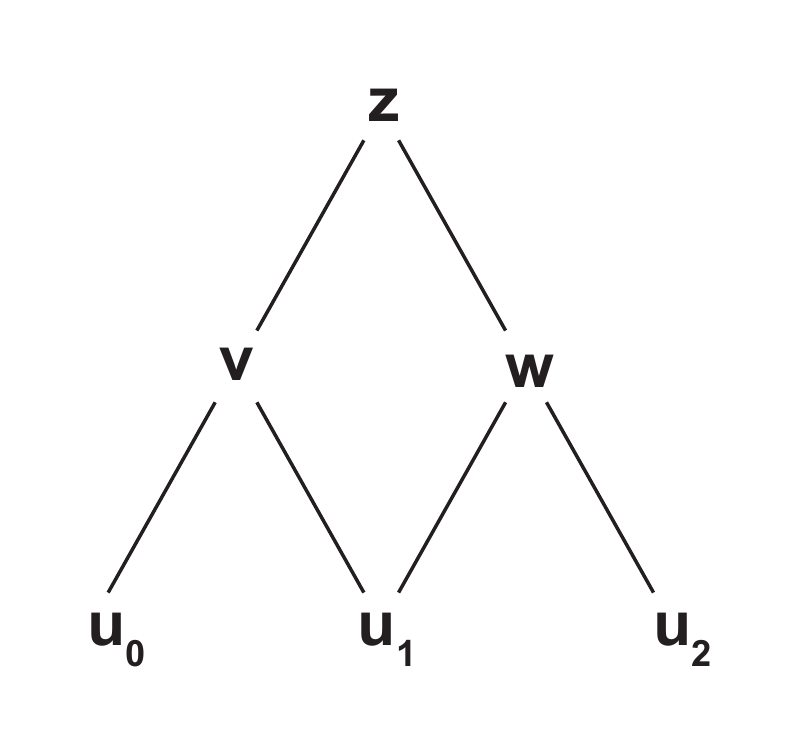}}
\caption{This figure illustrates the structure used in section \ref{sec:bi-quant-ids} to derive several identities involving bi-quantifications.}
\label{fig:inclusion-identity}
\end{figure}

\red
\subsection{The Chain Rule and the Direct Product Rule}
The chain rule (\ref{eq:chain-rule}) and the direct product rule (\ref{eq:direct-product-rule}) can be shown to be related by considering the product (joint) space \cite{GK:Symmetry}.  We consider a space $\mathbf{A}$ with elements $\ea_1$ and $\ea_2$ and a space $\mathbf{B}$ with elements $\eb_1$ and $\eb_2$.  By the direct product rule (\ref{eq:direct-product-rule}) we have that
\begin{multline}
b([(\ea_1 \vee \ea_2, \eb_1), (\ea_1 \vee \ea_2, \eb_1 \vee \eb_2)]) = \\
\qquad \qquad \qquad \qquad b([\ea_1 \vee \ea_2, \ea_1 \vee \ea_2]) b([\eb_1, \eb_1 \vee \eb_2]). \nonumber
\end{multline}
Since $b([\ea_1 \vee \ea_2, \ea_1 \vee \ea_2]) = 1$ from (\ref{eq:trivial-chain}), in the following section, we have that
\begin{equation}
b([(\ea_1 \vee \ea_2, \eb_1), (\ea_1 \vee \ea_2, \eb_1 \vee \eb_2)]) = b([\eb_1, \eb_1 \vee \eb_2]). \label{eq:chain-in-prod-space-1}
\end{equation}
Similarly, we can write
\begin{align}
b([(\ea_1, \eb_1), (\ea_1 \vee \ea_2, \eb_1)]) &= b([\ea_1, \ea_1 \vee \ea_2]) b([\eb_1, \eb_1]) \nonumber \\
&=  b([\ea_1, \ea_1 \vee \ea_2]),  \label{eq:chain-in-prod-space-2}
\end{align}
and
\begin{multline}
b([(\ea_1, \eb_1), (\ea_1 \vee \ea_2, \eb_1 \vee \eb_2)]) =\\
\qquad \qquad \qquad \qquad  b([\ea_1, \ea_1 \vee \ea_2]) b([\eb_1, \eb_1 \vee \eb_2]).  \label{eq:chain-in-prod-space-3}
\end{multline}

By writing $\ex = (\ea_1, \eb_1)$, $\ey = (\ea_1 \vee \ea_2, \eb_1)$, and $\ez = (\ea_1 \vee \ea_2, \eb_1 \vee \eb_2)$ we have that $\ex \leq \ey \leq \ez$ in the product (joint) space.  By substituting (\ref{eq:chain-in-prod-space-1}) and (\ref{eq:chain-in-prod-space-2}) into (\ref{eq:chain-in-prod-space-3}) we have the chain rule in the product space
\begin{equation}
b([\ex, \ez]) = b([\ex, \ey]) b([\ey, \ez]),
\end{equation}
and the chain rule and direct product rule are shown to be mutually consistent.
\black

\subsection{Bi-Quantification Identities} \label{sec:bi-quant-ids}
We conclude this section by deriving several identities.  Consider a chain where $\ex < \ey$.  The interval $[\ex, \ey]$ can be written as
$$
[\ex, \ey] = \big[[\ex,\ex], [\ex,\ey]\big]
$$
where $[\ex, \ex]$ is the trivial interval consisting of a single element.  Application of the chain rule implies that
\begin{equation}
b([\ex,\ey]) = b([\ex,\ex]) b([\ex,\ey]),
\end{equation}
so that, in general, we have that
\begin{equation} \label{eq:trivial-chain}
b([\ex,\ex]) = 1.
\end{equation}

Consider the structure in Figure \ref{fig:inclusion-identity} in which the elements $\eu_0$, $\eu_1$, and $\eu_2$ are mutually exclusive.  We consider intervals, such as $[\ez, \eu_0]$, for which $\eu_0 < \ez$.  Application of the sum rule yields
\begin{equation}
b([\ez, \eu_0]) = b([\eu_0, \eu_0]) + b([\ew, \eu_0]),
\end{equation}
where again, $b([\eu_0, \eu_0]) = 1$, and $\eu_0$ and $\ew$ are mutually exclusive.  However, we can also write
\begin{equation}
b([\ez, \eu_0]) = b([\eu_0, \eu_0]) + b([\eu_1, \eu_0]) + b([\eu_2, \eu_0]).
\end{equation}
Since the element $\eu_0$ is mutually exclusive to each of $\ew$, $\eu_1$, and $\eu_2$, their relationships must be quantified equally
\begin{equation}
b([\ew, \eu_0]) = b([\eu_1, \eu_0]) = b([\eu_2, \eu_0]) = 0,
\end{equation}
which implies that, in general, for mutually exclusive elements $\ea$ and $\eb$ we have that $b([\ea, \eb]) = 0$.
Moreover, for elements $\ey \geq \ex$, we have that $b([\ey, \ex]) = 1$.

\subsection{Bi-Quantifications: Product Rule} \label{sec:product-rule}
\begin{figure}
\centering
\makebox{\includegraphics[width=1\columnwidth]{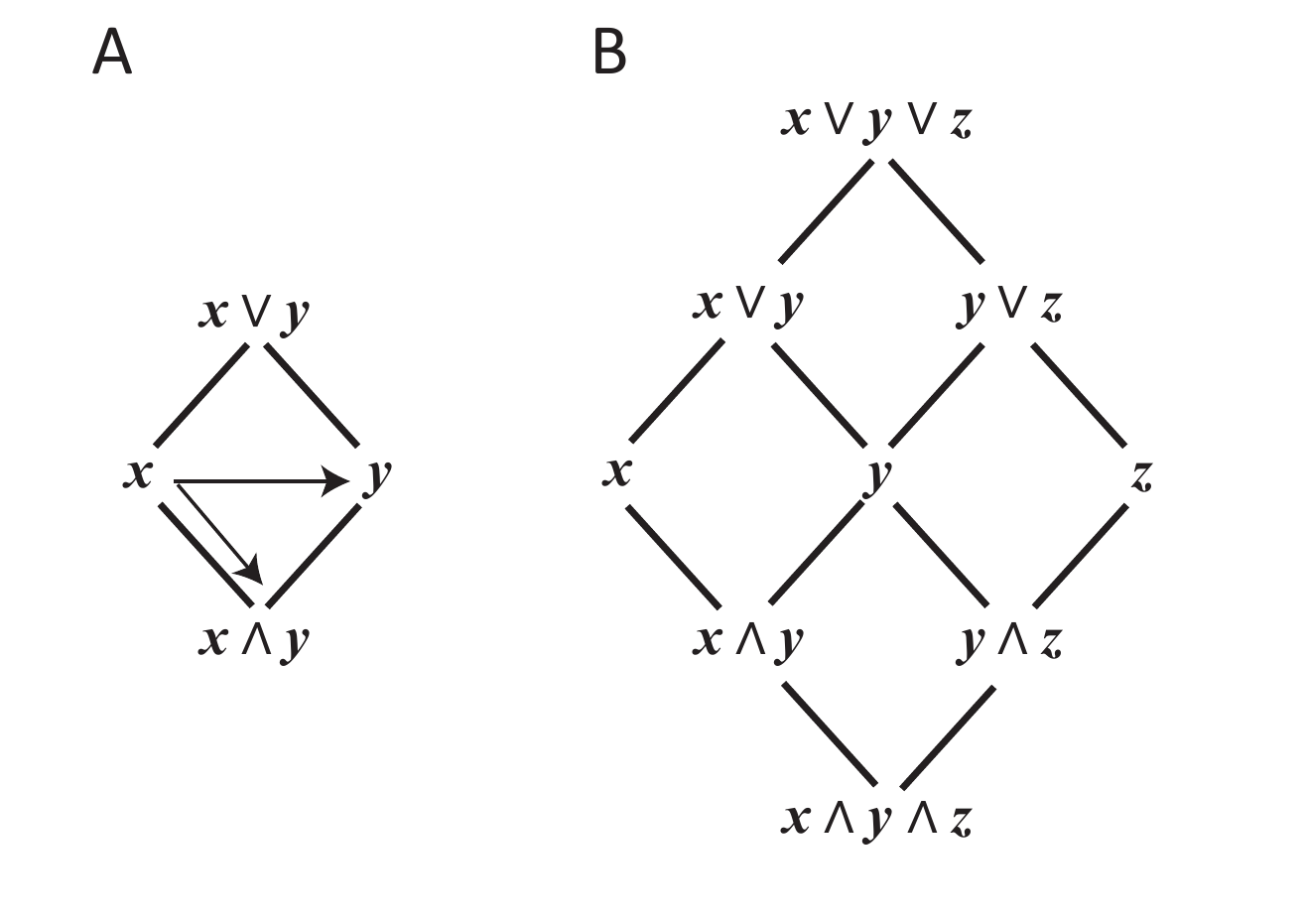}}
\caption{A. This lattice is used to demonstrate that $b([\ex \meet \ey, \ex]) = b([\ey,\ex])$.  B.  This larger lattice is used to derive the general product rule (\ref{eq:product-rule}). See the text in Section \ref{sec:product-rule} for details.} \label{fig:prod-rule}
\end{figure}

We now derive a more general product rule for bi-quantifications where the intervals do not necessarily comprise a chain.
Consider the lattice structure in Figure \ref{fig:prod-rule}A defined by $\ex$, $\ey$, $\ex \join \ey$ and $\ex \meet \ey$.  By considering the context to be $\ex$, the sum rule is
\begin{equation}
b([\ex,\ex]) + b([\ey,\ex]) = b([\ex \join \ey, \ex]) + b([\ex \meet \ey, \ex])
\end{equation}
Since $\ex \leq \ex$ and $\ex \leq \ex \join \ey$, we have that $b([\ex,\ex]) = b([\ex \join \ey, \ex])$ so that
\begin{equation} \label{eq:diamond-1}
b([\ex \meet \ey, \ex]) = b([\ey,\ex]).
\end{equation}

Consider the chain $\ex \meet \ey \meet \ez \leq \ex \meet \ey \leq \ex$ and the corresponding chain rule
\begin{equation} \label{eq:chain-prod}
b([\ex \meet \ey \meet \ez, \ex]) = b([\ex \meet \ey \meet \ez, \ex \meet \ey]) b([\ex \meet \ey, \ex]).
\end{equation}
By (\ref{eq:diamond-1}) the factor $b([\ex \meet \ey, \ex])$ on the right-hand side of (\ref{eq:chain-prod}) can be replaced by $b([\ey,\ex])$.
We now apply this technique to two other diamonds to replace the other two terms in (\ref{eq:chain-prod}).
Consider the diamond defined by the elements $\ex \meet \ey \meet \ez$, $\ex \meet \ey$, $\ey \join \ez$ and $\ez$ in the lattice in Figure \ref{fig:prod-rule}B.  This gives the relation
\begin{equation} \label{eq:diamond-2}
b([\ex \meet \ey \meet \ez, \ex \meet \ey]) = b([\ez, \ex \meet \ey])
\end{equation}
analogous to (\ref{eq:diamond-1}), which can be used to replace the first factor on the right-hand side of (\ref{eq:chain-prod}).
Last, considering the diamond defined by the elements $\ex$, $\ex \join \ey$, $\ey \meet \ez$, $\ex \meet \ey \meet \ez$ in the lattice in Figure \ref{fig:prod-rule}B, we find that
\begin{equation} \label{eq:diamond-3}
b([\ex \meet \ey \meet \ez, \ex]) = b([\ey \meet \ez, \ex]).
\end{equation}

Substituting (\ref{eq:diamond-1}), (\ref{eq:diamond-2}), and (\ref{eq:diamond-3}) into (\ref{eq:chain-prod}) we have
\begin{equation} \label{eq:product-rule}
b([\ey \meet \ez, \ex]) = b([\ez, \ex \meet \ey]) b([\ey, \ex]),
\end{equation}
which is the general \emph{product rule} for lattice elements.

\subsection{Bi-Quantifications: Bayes' Theorem}
With the product rule (\ref{eq:product-rule}) in hand, a Bayes' Theorem analogue is easily derived.  Commutativity of the $\meet$ operation implies that
\begin{align}
b([\ey \meet \ez, \ex]) &= b([\ez, \ex \meet \ey]) b([\ey, \ex]) \\
&= b([\ey, \ex \meet \ez]) b([\ez, \ex]).
\end{align}
Equating the two expressions on the right-hand side, and solving for $b([\ez, \ex \meet \ey])$, we have
\begin{equation}
b([\ez, \ex \meet \ey]) = \frac{b([\ey, \ex \meet \ez]) b([\ez, \ex])}{b([\ey, \ex])},
\end{equation}
which is the bi-quantification analogue of \emph{Bayes' Theorem}.

\subsection{Meaning and Bi-Quantifications}
It is commonplace to ascribe meaning, or a description, to a quantification.  In the case of bi-quantifications, some insight is gained by considering the \emph{zeta function}, which is used in order-theory to indicate
%
%
whether one element $\ex$ is included by another element $\ey$, as in $\ex \leq \ey$ \cite{Rota:foundations, Krishnamurthy:combinatorics}:
\begin{equation}
\zeta(\ex, \ey) = \begin{cases}
      0, & \text{if}\ \ex \nleq \ey \\
      1, & \text{if}\ \ex \leq \ey
    \end{cases}
    \qquad \text{(Zeta Function)}.
\end{equation}
As such, the zeta function serves to encode the order-theoretic structure.

The dual of the zeta function, defined by reversing the ordering relation,
\begin{equation}
\zeta^{\partial}(\ex, \ey) = \begin{cases}
      0, & \text{if}\ \ex \ngeq \ey \\
      1, & \text{if}\ \ex \geq \ey
    \end{cases}
\end{equation}
more closely mirrors the constraints derived for bi-quantifications since for elements $\ex \geq \ey$, we have that $b([\ex, \ey]) = 1$ and for mutually exclusive elements $\ea \meet \eb = \bot$, or $\ea \meet \eb = \varnothing$, we have that $b([\ea, \eb]) = 0$.  The major difference is that for elements $\ex \ngeq \ey$ that are not mutually exclusive, non-zero assignments are allowed.  In this sense, a bi-quantification is a generalization of order-theoretic inclusion ($\zeta^{\partial}(\ex, \ey)$) to a \emph{degree of inclusion}.  In this way, the meaning of a bi-quantification is inherited from the meaning of the ordering relation.

\begin{table*}[t]
\centering
  \begin{tabular}{ | l | l | }
    \hline
    Measures on Sets & $m(A \cup B) = m(A) + m(B) - m(A \cap B)$\\ \hline
    Probability Theory & $P(A \vee B|I) = P(A|I) + P(B|I) - P(A,B | I)$\\ \hline
    Information Theory & $I(A;B) = H(A) + H(B) - H(A,B)$\\ \hline
    Polya's Min Max Rule\cite{Polya&Szego} & $\max(A,B) = A + B - \min(A,B)$  \\ \hline
    Integral Divisors & $\log(\mbox{LCM}(A,B)) = \log(A) + \log(B) - \log(\mbox{GCD}(A,B))$ \\ \hline
    Euler Characteristic & $\chi = V - E + F$ \\ \hline
    Spherical Excess\cite{Klain&Rota} & $(A + B + C) - \pi$\\ \hline
    Three Slit Problem\cite{Sorkin:1994} & $I_3(A,B,C) = |A \sqcup B \sqcup C| - |A \sqcup B| - |A \sqcup C| - |B \sqcup C| + |A| + |B| + |C|$\\
    \hline
  \end{tabular}
  \caption{This table, \red which was first published in \cite{Knuth:FQXI2015}, \black provides some illustration of the ubiquity of the Sum Rule.  This rule is perhaps most familiar in the areas of measure theory on sets, where it is taken as a postulate, and probability theory.  However, it holds in numerous other situations where the concepts of closure, ordering, and associativity hold \cite{Knuth+Skilling:2012, Knuth:FQXI2015, Skilling+Knuth:MPQ}.  This is the reason that we add things when we combine them. \red (Here closure refers to the fact that when combining two things, the result is the same kind of thing.  For example, the union $\cup$ of two sets results in another set.) \black}
  \label{tab:sum-rule}
\end{table*}

\section{Summary} \label{sec:summary}
Lattices are partially ordered sets in which every pair of elements has a least upper bound called the join and a greatest lower bound called the meet.  Every lattice is an algebra where the join and meet are algebraic operations.  Lattice elements can be consistently quantified by a function $q$ that takes a lattice element to a real number: $q: \ex \in L \rightarrow q(\ex) \in \mathbb{R}$.  Fundamental properties, such as closure, associativity, and order,\cite{Skilling+Knuth:MPQ}, exhibited by lattices constrain quantification of the lattice elements.  Each symmetry leads to a constraint equation, which ensures that the symmetry is satisfied by the assigned quantifications.  These constraint equations are often referred to as rules or laws in specific applications.

The join of elements $\ex$ and $\ey$ of a lattice $L$ is quantified in accordance with the
\begin{align*}
&\mbox{\textbf{Sum Rule}} \\
&\qquad q(\ex \join \ey) = q(\ex) + q(\ey) - q(\ex \meet \ey).
\end{align*}
The fact that the sum rule is ubiquitous throughout the sciences, as illustrated in Table \ref{tab:sum-rule}, reflects the fact that it is founded on elementary symmetries that are easily satisfied \cite{Knuth+Skilling:2012, Knuth:FQXI2015, Skilling+Knuth:MPQ}.

Elements of the lattice product are quantified in accordance with the
\begin{align*}
&\mbox{\textbf{Direct Product Rule}} \\
&\qquad q\big( (\ea, \eb) \big) = q\big( \ea \big) q\big( \eb \big).
\end{align*}
\red This rule allows one to assign quantifications to joint spaces in a way that is consistent with the spaces considered individually.  Familiar applications include joint probabilities, as well as quantum amplitudes assigned to product spaces. \black

Directed intervals defined by two lattice elements are quantified by functions called bi-quantifications, in which the second argument is referred to as the context. Under the join of elements, directed intervals satisfy the
\begin{align*}
&\mbox{\textbf{Sum Rule}} \\
&\qquad b([\ex \join \ey, \ez]) = b([\ex,\ez]) + b([\ey,\ez]) - b([\ex \meet \ey,\ez]).
\end{align*}
Directed intervals of the lattice product are quantified in accordance with the
\begin{align*}
&\mbox{\textbf{Direct Product Rule}} \\
&\qquad b\big( [(\ea, \eb), (\ez,\et)] \big) = b\big( [\ea,\ez] \big) b\big( [\eb,\et] \big).
\end{align*}
Two directed intervals sharing a single element when chained together through concatenation are quantified by the
\begin{align*}
&\mbox{\textbf{Chain Rule}} \\
&\qquad b(\big[ [\ex, \ey], [\ey, \ez] \big]) = b([\ex,\ey]) b([\ey, \ez]).
\end{align*}
The chain rule can be \red expanded \black to relate intervals that are not necessarily chained resulting in the
\begin{align*}
&\mbox{\textbf{Product Rule}} \\
&\qquad b([ \ey \meet \ez, \ex ]) = b([\ez, \ex \meet \ey]) b([\ey, \ex])
\end{align*}
and an associated
\begin{align*}
&\mbox{\textbf{Bayes' Theorem}} \\
&\qquad b([\ez, \ex \meet \ey]) = \frac{b([\ey, \ex \meet \ez]) b([\ez, \ex])}{b([\ey, \ex])}.
\end{align*}
\red The chain rule, product rule and Bayes' Theorem are unique to bi-quantifications with the most familiar example being probability theory.  Although, the chain rule is also familiar the Feynman product rule for quantum amplitudes.\black

\red
By virtue of associativity and commutativity of the lattice join, the sum rule holds for \emph{all} lattices.  Similarly, associativity and commutativity of the lattice product implies that the direct product rule holds for \emph{all} lattice products.  Associativity of chaining implies that the chaining must be isomorphic to addition, and with the additional constraint of distributivity, we have that the chain rule, the product rule and Bayes' Theorem hold for \emph{bi-quantifications in all distributive lattices}.  While it is true that these results are more general since they hold for any system that satisfies the requisite symmetries, it should be noted that quantifications restricted to a particular range or otherwise constrained, as in valuations for which $x \leq y$ implies $q(x) \leq q(y)$, are not guaranteed to hold in general.

The co-information lattice is one example since it does not support non-negative valuations \cite{Bell:2003}.  Similarly, orthomodular lattices, relevant to quantum mechanics \cite{Birkhoff+vonNeumann:1936, vonNeumann:1996(1955), Varadarajan:1962, Gudder:1965, Gudder:1967, Holik:2013, Holik:2014, Holik:2016}, generally cannot support bounded positive valuations (measures) \cite{Greechie:1971}, referred to as states \cite{Bennett:1968}.  However, orthomodular lattices can support non-trivial bounded signed measures \cite{Greechie:1971}.  Despite the potential for specific lattice-dependent restrictions on the range of quantifications employed, the constraint equations derived in this paper hold for all lattices that exhibit the requisite symmetries.
\black

\red
\subsection{Examples and Applications}

\subsubsection{Probability Theory}
One of many applications of consistent quantification is that of probability theory where one focuses on bi-quantifications assigned to pairs of logical statements comprising a Boolean lattice ordered by logical implication.  The bi-quantification that we call probability inherits its meaning from the ordering relation so that probability represents the degree to which one logical statement implies another.  That is, \emph{probability is a degree of implication.}
Of course, the Boolean lattice need not be invoked as it is the symmetries of the Boolean algebra that constrain quantifications resulting in the sum and product rules discussed above \cite{Knuth+Skilling:2012, Skilling+Knuth:MPQ}.

\subsubsection{Questions, Entropy and Information}
Another application involves quantifying the degree to which one question answers another \cite{Cox:1979, Fry:Cybernetics, Knuth:Questions, Knuth:me2004, Knuth:AISTATS2005, Knuth:duality, Knuth:WCCI06, vanErp:2013, vanErp+etal:2017}.  By defining a question in terms of the set of all possible logical statements that answer it \cite{Cox:1979}, one can construct the lattice of questions \cite{Knuth:Questions} as a free distributive lattice \cite{Birkhoff:1967, Gratzer2003, Davey&Priestley}.  For example, consider a problem in which I have collected one piece of fruit that could be an apple, a banana, a cantaloupe, or a date.  The identity of the piece of fruit can be expressed with one of the following logical statements: $\ea = $ `It is an apple', $\eb = $ `It is a banana', $\ec = $ `It is a cantaloupe', or $\ed = $ `It is a date'.  The central issue $\eI$, which is the question that resolves the problem without ambiguity, is defined as the question that is answered only by one of the elements of the set
\begin{equation}
\eI = \{ \ea, \eb, \ec, \ed \}.
\end{equation}
The central issue can be phrased as $\eI = $ `Did you select an apple, a banana, a cantaloupe, or a date?',
and can be written as the lattice join (set union) of four elementary questions
\begin{align}
\eI &= \eA \vee \eB \vee \eC \vee \eD \\
&= \eA \cup \eB \cup \eC \cup \eD
\end{align}
in which $\eA = \{ \ea \}$, $\eB  = \{ \eb \}$, etc.

One could, of course, ask a less direct question, such as `Did you or did you not select an apple?'.  This question is answered by $\ea = $ `It is an apple' or any statement that implies that it is not an apple.  In short, this question is defined by the set of eight logical statements:
\begin{equation}
\eA \vee \eB\eC\eD = \{ \ea, \eb \vee \ec \vee \ed, \eb \vee \ec, \eb \vee \ed, \ec \vee \ed, \eb, \ec, \ed \}.
\end{equation}
Since a question is defined by all of the statements that answer it, the set of possible answers must include any statements that imply any statement in the set.  In lattice theory, such a set is known as a \emph{\textit{downset}}, a \emph{\textit{lower set}}, or an \emph{\textit{ideal}} or \emph{\textit{order ideal}}.  Questions, defined as downsets of answers, are naturally ordered by subset inclusion.  For example, since
$$\eI \subseteq \eA \vee \eB\eC\eD,$$
we say that the question $\eI$ answers the question $\eA \vee \eB\eC\eD$ and we write
$$\eI \leq \eA \vee \eB\eC\eD.$$
Questions ordered by subset inclusion are then naturally ordered based on which questions answer, or resolve, others.  This is the basic order-theoretic structure.

One can quantify the degree to which one question answers another by employing a bi-quantification.  For example, one can express the degree to which the question $\eA \vee \eB\eC\eD = \mbox{`Did you or did you not select an apple?'}$ resolves the central issue $\eI$ with the bi-quantification $b(\eA \vee \eB\eC\eD, \eI)$.  We have previously shown \cite{Knuth:me2004, Knuth:duality, Knuth:WCCI06} that consistency with the probabilities of the logical statements requires that the bi-quantification of a question that partitions the set of possible answers is proportional to the Shannon entropy \cite{Shannon&Weaver} with
\begin{equation}
b(\eA \vee \eB\eC\eD, \eI) = C H(\ea, \eb \vee \ec \vee \ed),
\end{equation}
in which $C$ is a normalization constant and $H(\ea, \eb \vee \ec \vee \ed)$ is the Shannon entropy given by
\begin{align}
H(\ea, \eb \vee \ec \vee \ed) = &- \Bigl( Pr(\ea | \ei) \log_2 \left( Pr(\ea | \ei) \right) + \\
& \bigl( Pr(\eb \vee \ec \vee \ed | \ei) \bigr) \log_2 \bigl( Pr(\eb \vee \ec \vee \ed | \ei) \bigr) \Bigr), \nonumber
\end{align}
where $\ei = \ea \vee \eb \vee \ec \vee \ed$ is the truism for the hypothesis space, and
\begin{equation}
Pr(\eb \vee \ec \vee \ed | \ei) = Pr(\eb | \ei) + Pr(\ec | \ei) + Pr(\ed | \ei).
\end{equation}

Furthermore, we can write
\begin{equation}
b(\eI, \eI) = C H(\ea, \eb, \ec, \ed),
\end{equation}
for which
\begin{align}
H(\ea, \eb, \ec \ed) = - \Bigl( & Pr(\ea | \ei) \log_2 \bigl( Pr(\ea | \ei) \bigr) + \nonumber \\
& Pr(\eb | \ei) \log_2 \bigl( Pr(\eb | \ei) \bigr) + \nonumber \\
& Pr(\ec | \ei) \log_2 \bigl( Pr(\ec | \ei) \bigr) + \nonumber \\
& Pr(\ed | \ei) \log_2 \bigl( Pr(\ed | \ei) \bigr) \Bigr).
\end{align}
The fact that the bi-quantification $b(\eI, \eI) = 1$ allows us to find the normalization constant $C$:
\begin{align}
C^{-1} = H(\ea, \eb, \ec) = - \Bigl( & Pr(\ea | \ei) \log_2 \bigl( Pr(\ea | \ei) \bigr) + \nonumber \\
& Pr(\eb | \ei) \log_2 \bigl( Pr(\eb | \ei) \bigr) + \nonumber \\
& Pr(\ec | \ei) \log_2 \bigl( Pr(\ec | \ei) \bigr) + \nonumber \\
& Pr(\ed | \ei) \log_2 \bigl( Pr(\ed | \ei) \bigr) \Bigr).
\end{align}
So that these bi-quantifications, such as $b(\eA \vee \eB\eC\eD, \eI)$, which quantifies the \emph{\textit{relevance}} of the question $\eA \vee \eB\eC\eD$ to the central issue $\eI$ are ratios of entropies
\begin{equation}
b(\eA \vee \eB\eC\eD, \eI) = \frac{H(\ea, \eb \vee \ec \vee \ed)}{H(\ea, \eb, \ec, \ed)},
\end{equation}
just as probabilities are ratios of measures \cite{Knuth+Skilling:2012, Skilling+Knuth:MPQ}.
As a result, the degree to which the question $\eA \vee \eB\eC\eD = \mbox{`Did you or did you not select an apple?'}$ resolves the issue $\eI = $ `Did you select an apple, a banana, a cantaloupe, or a date?', denoted $b(\eA \vee \eB\eC\eD, \eI)$ depends on the probabilities of the possible answers via Shannon's entropy.  Furthermore, the relevance is bounded
\begin{equation}
0 \leq b(\eA \vee \eB\eC\eD, \eI) \leq 1,
\end{equation}
with $b(\eA \vee \eB\eC\eD, \eI) = 0$ indicating that the question is not relevant because it is already known that an apple is not selected ($Pr(\ea | \ei) = 0$), and the limit with $b(\eA \vee \eB\eC\eD, \eI) \simeq 1$ indicating that the question resolves the issue with near certainty.

We can now consider the lattice join of two questions, which is defined by their set union.  Our earlier results, obtained for lattices in general indicate that we will have a sum rule.  The question $(\eA\eB \vee \eB\eC\eD)$ can be written as the join of two questions
\begin{align}
(\eA\eB \vee \eB\eC\eD) &= (\eA\eB \vee \eC\eD) \vee (\eA \vee \eB\eC\eD) \\
&= \{ \ea \vee \eb, \ea, \eb, \ec \vee \ed, \ec, \ed \} \cup \nonumber  \\
& \qquad \{ \ea, \eb \vee \ec \vee \ed, \eb \vee \ec, \eb \vee \ed, \ec \vee \ed, \eb, \ec, \ed \} \nonumber  \\
&= \{ \ea \vee \eb, \ea, \eb, \eb \vee \ec \vee \ed, \eb \vee \ec, \eb \vee \ed, \ec \vee \ed, \eb, \ec, \ed \}, \nonumber
\end{align}
whereas their meet is
\begin{align}
(\eA\eB \vee \eC\eD) \wedge (\eA \vee \eB\eC\eD) &= \{ \ea \vee \eb, \ea, \eb, \ec \vee \ed, \ec, \ed \} \cap \nonumber  \\
 & \{ \ea, \eb \vee \ec \vee \ed, \eb \vee \ec, \eb \vee \ed, \ec \vee \ed, \eb, \ec, \ed \} \nonumber \\
&= \{ \ea, \eb, \ec \vee \ed, \ec, \ed \} \nonumber  \\
&= (\eA \vee \eB \vee \eC\eD).
\end{align}
The sum rule can be used to compute the relevance of $(\eA\eB \vee \eB\eC\eD)$ to the central issue $\eI$ in terms of the following relevances
\begin{align}
& b\left( (\eA\eB \vee \eB\eC\eD), \eI \right) \\
&\qquad = b\left( (\eA\eB \vee \eC\eD), \eI \right) + b\left( (\eA \vee \eB\eC\eD), \eI \right) - b\left( (\eA \vee \eB \vee \eC\eD), \eI \right) \nonumber \\
&\qquad = \frac{H(\ea \vee \eb, \ec \vee \ed)}{H(\ea, \eb, \ec, \ed)} + \frac{H(\ea, \eb \vee \ec \vee \ed)}{H(\ea, \eb, \ec, \ed)} - \frac{H(\ea, \eb, \ec \vee \ed)}{H(\ea, \eb, \ec, \ed)}. \nonumber
\end{align}
Due to the exclusion term that is subtracted, the relevance of a question that does not partition the answers, such as this one, can be negative \cite{Bell:2003, Knuth:duality, vanErp+etal:2017}.  However, restricting oneself to partition questions, which partition the top answers, such as $(\eA\eB \vee \eC\eD)$, $(\eA \vee \eB\eC\eD)$, and $\eI = (\eA \vee \eB \vee \eC \vee \eD)$, one is left with a bi-valuation that has non-negative relevance values.

The sum rule in the context of questions also results in the familiar mutual information relation
\begin{equation}
MI(\eX, \eY) = H(\eX) + H(\eY) - H(\eX, \eY).
\end{equation}

The chain rule (product rule) is useful when changing context.  For example, since $\eI = (\eA \vee \eB \vee \eC \vee \eD) \leq (\eA \vee \eB \vee \eC\eD) \leq (\eA \vee \eB\eC\eD)$, we can write
\begin{multline}
b\left( (\eA \vee \eB\eC\eD), \eI \right) = \\
b\left( (\eA \vee \eB\eC\eD), (\eA \vee \eB \vee \eC\eD) \right) b\left( (\eA \vee \eB \vee \eC\eD), \eI \right),
\end{multline}
which can be used to determine $b\left( (\eA \vee \eB\eC\eD), (\eA \vee \eB \vee \eC\eD) \right)$
\begin{align}
&b\left( (\eA \vee \eB\eC\eD), (\eA \vee \eB \vee \eC\eD) \right) = \frac{b\left( (\eA \vee \eB\eC\eD), \eI \right)}{b\left( (\eA \vee \eB \vee \eC\eD), \eI \right)} \nonumber \\
&\qquad \qquad = \frac{H(\ea, \eb \vee \ec \vee \ed)}{H(\ea,\eb,\ec,\ed)} \frac{H(\ea,\eb,\ec,\ed)}{H(\ea, \eb, \ec \vee \ed)} \nonumber \\
&\qquad \qquad = \frac{H(\ea, \eb \vee \ec \vee \ed)}{H(\ea, \eb, \ec \vee \ed)},
\end{align}
so that the relevance relationship among less precise questions is again quantified as a ratio of entropies.

This is a significant result in that it indicates that the domain of application of Shannon's entropy is not limited to the communication channels for which it was originally derived \cite{Shannon&Weaver}.  Here we see that Shannon's entropy quantifies the degree to which some questions answer other questions, which explains the wide applicability of the measure.

In addition to casting information theory in a new light, the ability to quantify relevance among questions would have a significant impact on artificial intelligence systems both in the context of human interaction \cite{Knuth:PhilTrans, Knuth:AISTATS2005} and in the area of autonomous experimental design \cite{Knuth+Erner+Frasso:2007, Knuth+Center:2008, Knuth+Center:2010}.

\subsubsection{Additional Applications: Concept Lattices and Quantum Mechanics}
Since these rules for the consistent quantification of lattices widely apply, it is to be expected that there will be other applications.  For example, such rules would allow for the consistent quantification of concept lattices \cite{Wille:1992, Godin+etal:1995, Buelohlavek:2004, Yao:2004} in computer science, thus extending the concept lattice algebra to a calculus.

The appreciation that the symmetries are what is critical in constraining the sum and product rules enables these ideas to be applied to problems that exhibit those symmetries.  For example, we have applied the same concepts of consistent quantification to quantum measurement sequences and by assuming quantifications based on pairs of numbers we have derived the complex sum and product rules \cite{GKS:PRA, GK:Symmetry, Skilling+Knuth:MPQ} for manipulating Feynman amplitudes.  There are efforts underway by Holik and colleagues to apply this approach to the non-distributive lattice of subspaces of the Hilbert space in quantum mechanics \cite{Holik+etal:2013, Holik:2014, Holik+etal:2015}.

With this methodology firmly in place, it will be interesting to discover what additional applications await.
\black

\bibliographystyle{amsplain}
\bibliography{C:/Users/Kevinator/kevin/papers/bib/knuth53}

\end{document}